\documentclass[iop]{emulateapj}
\usepackage{graphicx,amsmath,amsthm,amssymb,multirow,color}

\definecolor{Mygrey}{gray}{0.75}

\shorttitle{Discovery of CRRLs in M82}
\shortauthors{Morabito et al.}

\begin{document}

\title{Discovery of Carbon Radio Recombination Lines in M82}
  
\author{Leah K. Morabito\altaffilmark{1}, J.~B.~R. Oonk\altaffilmark{1,2}, Francisco Salgado\altaffilmark{1}, M.~Carmen Toribio\altaffilmark{2}, H.~J.~A. R\"{o}ttgering\altaffilmark{1}, A.~G.~G.~M. Tielens\altaffilmark{1}, Rainer Beck\altaffilmark{3}, Bj\"{o}rn Adebahr\altaffilmark{3}, Philip Best\altaffilmark{4}, Robert Beswick\altaffilmark{5}, Annalisa Bonafede\altaffilmark{6}, Gianfranco Brunetti\altaffilmark{7}, Marcus Br\"{u}ggen\altaffilmark{6}, Krzysztof T. Chy\.zy\altaffilmark{8}, J.~E. Conway\altaffilmark{9}, Wim van Driel\altaffilmark{10}, Jonathan Gregson\altaffilmark{11}, Marijke Haverkorn\altaffilmark{12}$^,$\altaffilmark{1}, George Heald\altaffilmark{2}$^,$\altaffilmark{13}, Cathy Horellou\altaffilmark{9}, Andreas~Horneffer\altaffilmark{3}, Marco Iacobelli\altaffilmark{2}, Matt J. Jarvis\altaffilmark{14}$^,$\altaffilmark{15}, Ivan Marti-Vidal\altaffilmark{9}, George Miley\altaffilmark{1}, D.~D. Mulcahy\altaffilmark{16}, Emanuela Orr\'{u}\altaffilmark{2}$^,$\altaffilmark{12}, Roberto Pizzo\altaffilmark{2}, A.~M.~M. Scaife\altaffilmark{16}, Eskil Varenius\altaffilmark{9}, Reinout J. van Weeren\altaffilmark{17}, Glenn J. White\altaffilmark{11}$^,$\altaffilmark{18}, Michael W. Wise\altaffilmark{2} }

\affil{\vspace{0pt}\\ $^{1}$Leiden Observatory, Leiden University, P.O. Box 9513, 2300 RA, Leiden, The Netherlands \\
$^{2}$Netherlands Institute for Radio Astronomy (ASTRON), Postbus 2, 7990 AA Dwingeloo, The Netherlands \\
$^{3}$MPI f\"{u}r Radioastronomie, Auf dem H\"{u}gel 69, 53121 Bonn, Germany \\
$^{4}$SUPA, Institute for Astronomy, Royal Observatory Edinburgh, Blackford Hill, Edinburgh, EH9 3HJ, UK \\
$^{5}$Jodrell Bank Centre for Astrophysics, University of Manchester, Oxford Road, Manchester, M13 9PL, UK \\
$^{6}$Hamburg Observatory, University of Hamburg, Gojenbergsweg 112, 21029 Hamburg, Germany \\
$^{7}$INAF-Istituto di Radioastronomia, via P. Gobetti 101, 40129 Bologna, Italy \\
$^{8}$Obserwatorium Astronomiczne Uniwersytetu Jagiello\'nskiego, ul. Orla 171, 30-244 Krak\'ow, Poland \\
$^{9}$Department of Earth and Space Sciences, Chalmers University of Technology, Onsala Space Observatory, SE-439 92 Onsala, Sweden \\
$^{10}$GEPI, Observatoire de Paris, UMR 8111, CNRS, Universit\'{e} Paris Diderot, 5 place Jules Janssen, 92190 Meudon, France \\
$^{11}$Department of Physical Sciences, The Open University, Milton Keynes MK7 6AA, England \\
$^{12}$Department of Astrophysics/IMAPP, Radboud University Nijmegen, P.O. Box 9010, 6500 GL Nijmegen, The Netherlands \\
$^{13}$Kapteyn Astronomical Institute, University of Groningen, PO Box 800,
9700 AV, Groningen, The Netherlands \\
$^{14}$Astrophysics, Department of Physics, Keble Road, Oxford OX1 3RH, UK \\
$^{15}$Department of Physics, University of the Western Cape, Private Bag X17, Bellville 7535, South Africa \\
$^{16}$School of Physics \& Astronomy, University of Southampton, Highfield, Southampton, SO17 1BJ, UK \\ 
$^{17}$Harvard-Smithsonian Center for Astrophysics, 60 Garden Street, Cambridge, MA 02138, USA \\
$^{18}$RALSpace, The Rutherford Appletion Laboratory, Chilton, Didcot, Oxfordshire OX11 0NL, England}

\begin{abstract}
Carbon radio recombination lines (RRLs) at low frequencies ($\lesssim500\,\textrm{MHz}$) trace the cold, diffuse phase of the interstellar medium, which is otherwise difficult to observe. 
We present the detection of carbon RRLs in absorption in M82 with LOFAR in the frequency range of $48-64\,\textrm{MHz}$. 
This is the first extragalactic detection of RRLs from a species other than hydrogen, and below $1\,\textrm{GHz}$. 
Since the carbon RRLs are not detected individually, we cross-correlated the observed spectrum with a template spectrum of carbon RRLs to determine a radial velocity of $219\pm9\,\textrm{km s}^{-1}$. 
Using this radial velocity, we stack 22 carbon-$\alpha$ transitions from quantum levels $n=468-508$ to achieve an $8.5\sigma$ detection. 
The absorption line profile exhibits a narrow feature with peak optical depth of $3\times10^{-3}$ and FWHM of $31\,\textrm{km s}^{-1}$. 
Closer inspection suggests that the narrow feature is superimposed on a broad, shallow component. 
The total line profile appears to be correlated with the 21 cm \ion{H}{1} line profile reconstructed from \ion{H}{1} absorption in the direction of supernova remnants in the nucleus. 
The narrow width and centroid velocity of the feature suggests that it is associated with the nuclear starburst region. 
It is therefore likely that the carbon RRLs are associated with cold atomic gas in the direction of the nucleus of M82.
\end{abstract}

\keywords{galaxies: ISM --- galaxies: individual(\objectname{M 82}) --- ISM: general --- radio lines: ISM --- radio lines: galaxies}

\section{Introduction}
The nearby \citep[$3.52 \pm 0.02\textrm{ Mpc}$;][]{jacobs09} nuclear starburst galaxy M82 has been observed to host a wide range of phases of the interstellar medium (ISM). 
Observations of disrupted \ion{H}{1} within the disk show that the neutral gas is more concentrated in the nuclear region \citep{yun93}. There is a rotating ring of molecular gas in the nucleus, as seen from observations of HCN, HCO+, CO(2-1) \citep{kepley14}, OH \citep[e.g.,][]{argo10}, and CO images \citep[e.g.,][]{westmoquette13}. 
Numerous \ion{H}{2} regions are seen \citep[e.g.,][]{mcd02,gandhi11} in this area.  
The nuclear region is also studded with compact, bright supernova remnants  \citep[SNRs; e.g.,][]{muxlow94,fenech10}. The spectral turnovers of the SNRs \citep[Varenius et al., in prep.,][]{wills97} as well as the overall spectrum \citep[Varenius et al., in prep.,][]{adebahr13} indicate the presence of free-free absorption by ionized gas. 
The complex interplay of all these components of the ISM is not fully understood.

Carbon radio recombination lines can help characterize the cold, diffuse phase of the interstellar medium (ISM). 
When free electrons recombine with atoms at quantum numbers $n\gtrsim 50$, the decreased energy spacing of subsequent levels produces radio recombination lines (RRLs). 
By comparing observations of RRLs with detailed physical models, we can determine information on the physical properties, such as electron temperature and density, of the gas in which the RRLs originate \citep[e.g.,~Salgado et al.,~in preparation;][]{ww82,shaver75,dupree69}. 
At frequencies $\lesssim500\,\textrm{MHz}$, RRLs are spaced closely enough that wide-bandwidth instruments like the Low Frequency Array \citep[LOFAR;][]{vh13} are able to track the dependence of line properties on quantum number within a single observation. 
These properties make RRLs a powerful tool for determining the temperature and density of their host phases of the ISM. 

RRLs fall into two categories: discrete and diffuse. 
Discrete RRLs trace warm ($T_e\sim10^4\textrm{ K}$), high-density ($n_e > 100\textrm{ cm}^{-3}$) gas associated with \ion{H}{2} regions \citep{palmer67}. They are predominantly seen at frequencies above $\sim 1\textrm{ GHz}$ and originate from hydrogen, helium, and carbon \citep[e.g.,][]{poppi07,ks05,rg91}. These types of RRLs have been detected in a handful of nearby bright star-forming galaxies \citep[e.g.,][]{shaver77,anath93,rr04,roy08}. 

Diffuse carbon RRLs (CRRLs) trace the cold neutral medium (CNM), which is cold ($T_e\sim100\textrm{ K}$), and diffuse ($n_e \lesssim 0.1\textrm{ cm}^{-3}$). Diffuse RRLs are observed at frequencies below $1\textrm{ GHz}$. 
Typically the CNM has ionization levels that are too low to produce hydrogen and helium lines, and only CRRLs are observed. 
The ionization energy for atomic carbon is only $11.3\,\textrm{eV}$, lower than that of hydrogen, $13.6\,\textrm{eV}$. Some photons that can escape \ion{H}{2} regions can therefore ionize carbon, and thus \ion{C}{2} is expected to be the dominant state of carbon in the ISM. While discrete RRLs are used extensively to study star forming regions \citep[e.g.,][]{anderson11,rgm92}, not much is yet known about the CNM associated with diffuse RRLs, even in our Galaxy. Low frequency observations have shown that the CRRL emitting and absorbing gas is prevalent on scales of degrees along the Galactic plane \citep{ema95,ka01}. 
Pinhole studies in the direction of \ion{H}{2} regions \citep[e.g.,][]{gk91}, supernova remnants \citep[Cassiopeia A; e.g.,][]{asgekar13,pae89,ks81} or bright background extragalactic sources \citep{oonk14} have detected CRRLs on smaller scales. 
The new Low Frequency Array \citep[LOFAR;][]{vh13} has observed CRRLs in our Galaxy.  Over the next several years, LOFAR will perform a CRRL survey of the Galactic plane, on scales from degrees down to several arcseconds, producing maps that will provide a comprehensive picture of the Galactic CNM by quantifying the average gas temperatures and densities, as well as abundances.

In this paper, we present the first detection of extragalactic CRRLs, observed in M82 at frequencies near $60\textrm{ MHz}$ with LOFAR. 
This is the first extragalactic detection of diffuse RRLs, and the first extragalactic detection of RRLs from a species other than hydrogen. 
This detection opens up the possibility of tracing the evolution of the CNM through all stages of galaxy formation.

Section~\ref{sec:dr} outlines observations, data reduction, and imaging. Section~\ref{sec:spec} describes the extraction and processing of the spectra, cross-correlation of the overall spectrum to determine a velocity, and subband stacking to achieve a detection. Results are presented in \S$\,$4. Discussion and conclusions follow in \S$\,$\ref{sec:diss} and \S$\,$\ref{sec:conclusions}, respectively. 

\section{Observations and Data Reduction}
\label{sec:dr}
M82 was observed with a total on-source time of 5.0 hours on 21 February 2013 with the LOFAR low-band antenna (LBA; $10-80\,\textrm{MHz}$) stations, as part of early science (Cycle 0) data obtained for the LOFAR Survey of nearby galaxies (LC0\_043, PI: R. Beck). 
We used 13 remote and 24 core stations for this observation, giving baselines between 90~m and 85~km.
Each LBA station contains 96 dual polarization dipole antennas, 48 of which can be used simultaneously, and has a total radius of $87\,\textrm{m}$. 
We used the LBA\_OUTER configuration, where the outermost more sparsely spaced 48 antennas in each station record data. 
This configuration has a reduced field of view compared to other configurations, and mitigates the problem of mutual coupling between closely spaced antennas, providing slightly increased sensitivity. 

Each dual-polarization dipole antenna provides four linear correlation products which were used to reconstruct Stokes I images.
We obtained complete frequency coverage between 30 and 78~MHz with 1~s time resolution. 
This bandwidth was divided in 244 subbands, each 0.1953~MHz wide, and further subdivided into 128 channels, resulting in channel widths ranging from 6\,--\,16~km~s$^{-1}$. Using the total available bandwidth of 488 subbands and the multibeam capability of LOFAR we simultaneously observed the calibrator source 3C196 with identical frequency coverage.

The calibrator data were processed by the observatory pipeline, starting by flagging radio frequency interference (RFI) with the AOFlagger \citep{offringa10}. 
Typically, a few percent of the data were flagged due to RFI, consistent with a study of the LOFAR RFI environment \citep{offringa13}. 
The calibrator data were then averaged to 16 channels and 6 seconds before calibration with the BlackBoard Selfcal software system \citep{pandey09}. 
The target data were flagged with the AOFlagger, averaged to 6 seconds and 32 channels and the amplitude calibration obtained from 3C196 was applied. 
The velocity sampling of the individual subbands in target data after averaging ranges from $29-37\textrm{ km s}^{-1}$.  

For imaging we selected only the 24 LOFAR core stations, which all observe with the same clock, making correction for time delays unnecessary. 
The proximity of the stations to each other also mitigates the chance of observing the target through severely different ionospheric conditions.
Image cubes of a region with an area of $6\times6$ degree$^2$ centred on M82 were made with AWimager \citep{tasse13}, imaging and cleaning each channel individually. 
We used Briggs weighting \citep{briggs95} with a robust value of 0.0 to create images with resolutions ranging between $259\times340\,$arcsec$^{2}$ and $322\times425\,$arcsec$^{2}$. 
At this resolution M82 is unresolved.
Low resolution imaging parameters were deliberately chosen to speed up the computationally expensive imaging process. LOFAR does not support Doppler tracking so we have Doppler-corrected the data post imaging. In total we imaged 69 subbands in the range 50-64 MHz, centred on the peak response of the LBA \citep{vh13}. 

\section{Spectral Processing}
\label{sec:spec}

\subsection{Individual subband processing}
\label{sec:spec1}
\begin{figure}
\begin{center}
\includegraphics[width=0.2\textwidth, clip, trim=0.5cm 0.9cm 1.8cm 0cm]{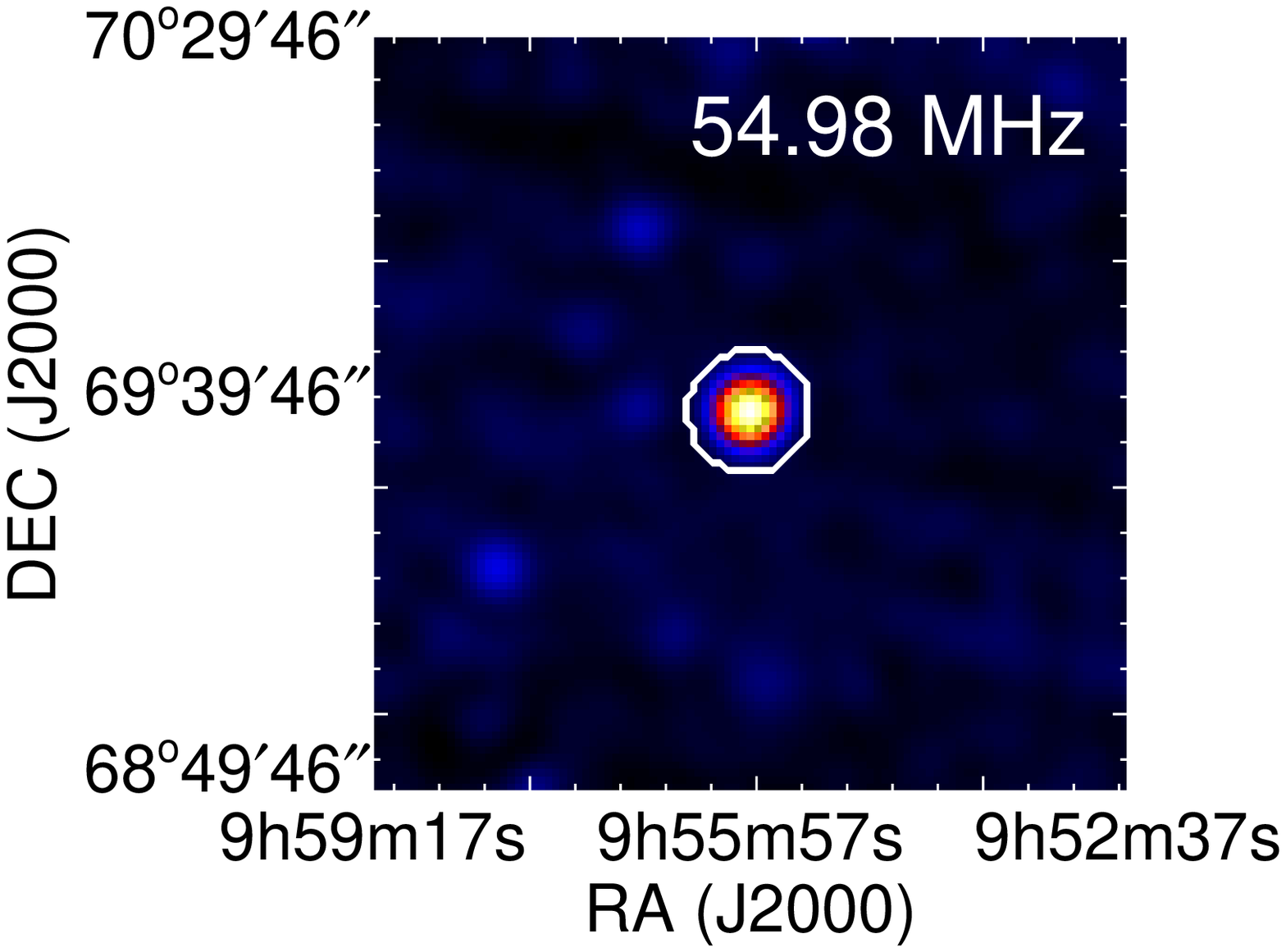}\hspace{-15pt}
\includegraphics[width=0.15\textwidth, clip, trim=2.1cm 0.9cm 3.1cm 0cm]{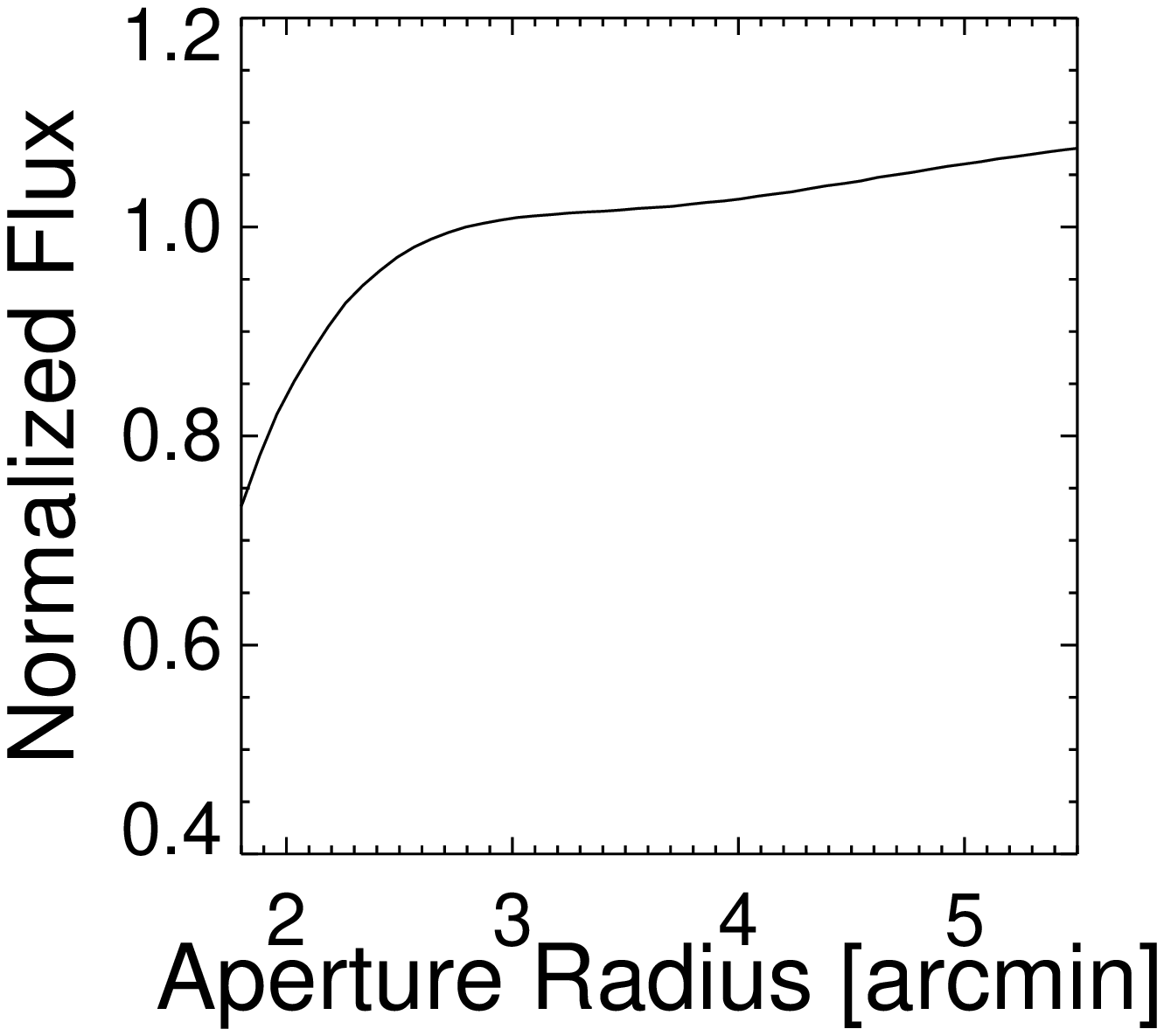}\hspace{-7pt}
\includegraphics[width=0.15\textwidth, clip, trim=2.1cm 0.9cm 3.1cm 0cm]{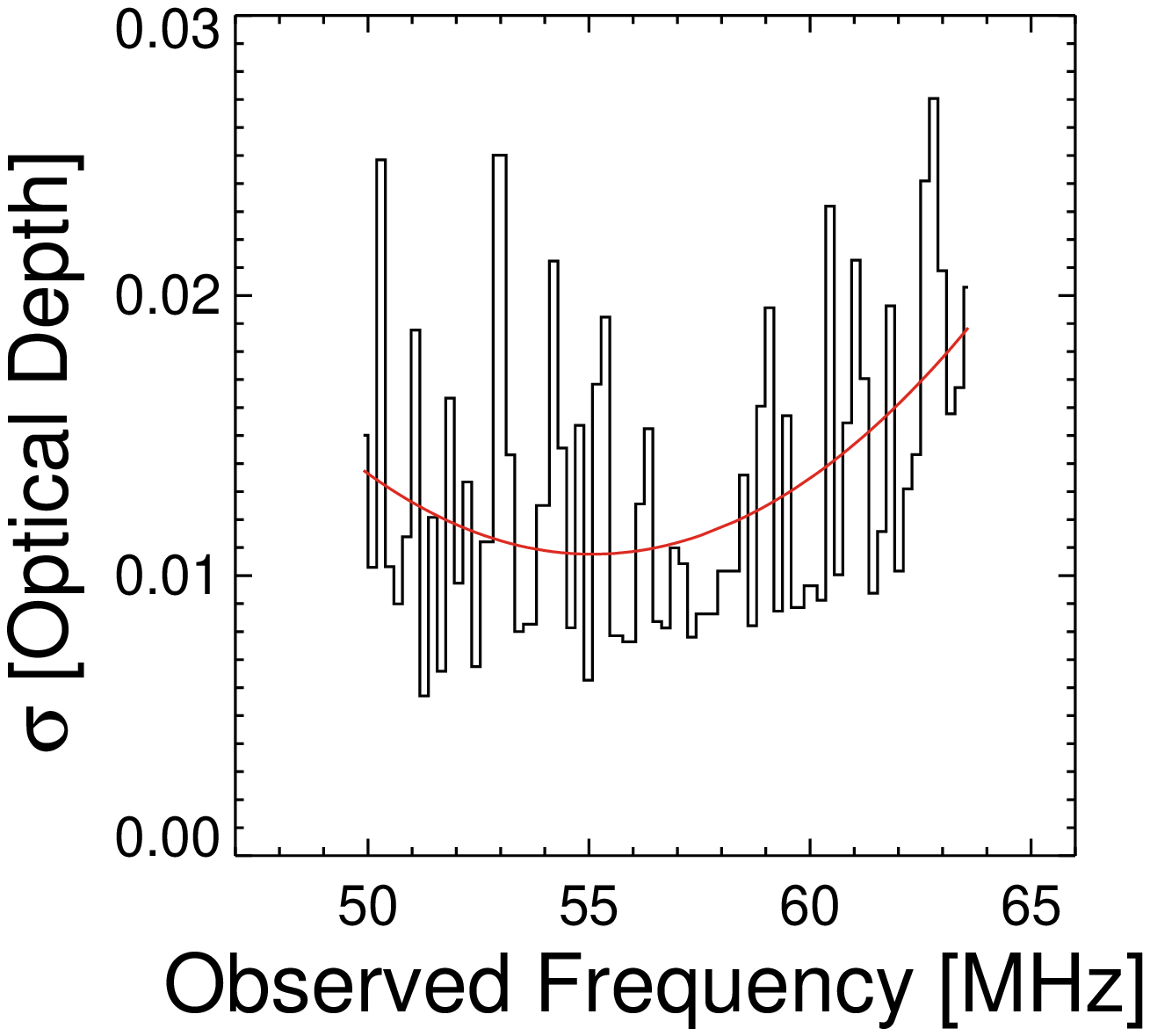} \\
\includegraphics[width=0.5\textwidth, clip, trim=0.2cm 1cm 1.9cm 6.5cm]{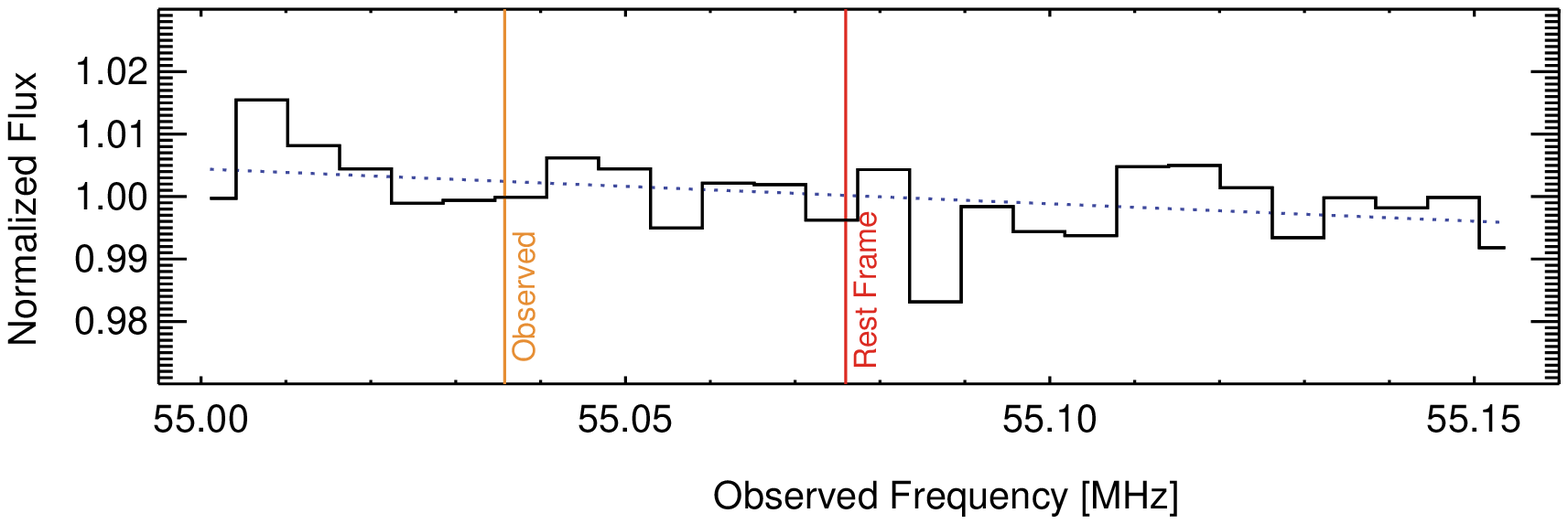}
\caption{A demonstration of the spectral extraction process. The top left panel shows the centre of an image obtained by averaging all channels maps within one subband, with a white contour marking the extraction aperture (diameter $\sim15$ arcmin). The top centre panel shows the curve-of-growth for the same subband used to determine the size of the aperture. The top right panel shows the optical depth standard deviation as a function of subband (frequency), with excessively noisy subbands filtered out. The bottom panel displays the normalised raw extracted spectrum for the same subband as the top left and center panels. A linear fit is overplotted as a blue dotted line, and observed and rest frame ($z=0.00073$) frequencies of the $\alpha$-transition CRRL in this subband are labelled. \label{fig:f1}}
\end{center}
\end{figure}

We extracted the spatially integrated spectra from elliptical apertures, where the appropriate elliptical dimensions in each subband were determined by fitting a 2D Gaussian to the point source in the centre and using the curve-of-growth (see Fig.~\ref{fig:f1}) to select an aperture size that captured the total flux density of the point source but did not include nearby background sources.
We calculated the standard deviation for each subband and implemented automatic removal of subbands with excessively large standard deviations ($\geq 10\times \sigma_{\textrm{median}}$), verified by visual inspection.
Five subbands were removed from further processing. We inspected each subband for individual detections of lines, which were not seen. 
In order to increase the signal-to-noise ratio, we stacked CRRL $\alpha$-transitions ($\Delta n=1$, for $468 < n < 508$) to form an average line profile.

\subsection{Measuring the velocity/redshift}
Maximizing the signal-to-noise ratio of a stacked line profile requires precisely stacking the centres of each individual line, which requires an accurate radial velocity (redshift). 
While the systemic velocity of M82 has been measured as $210\pm20\,\textrm{km s}^{-1}$ relative to the local standard of rest (LSR), observations show that even in the nuclear region there is a spread of $\sim200\,\textrm{km s}^{-1}$ in velocity \citep[e.g.,][]{wills98,kepley14}. 
We constrained the velocity for the CRRL absorbing gas by cross-correlating a spectral template of CRRLs (using Gaussian line profiles at known CRRL frequencies) with the entire observed spectrum over a range of redshifts. 
The template line peaks were normalised to a value of $3\times10^{-3}$. 
For the less certain line width parameter, we ran a series of models with different line widths, starting at $v=15\,\textrm{km s}^{-1}$ (to avoid cross-correlating with noise peaks). 
The magnitude of the cross-correlation peaks at $v=210\,\textrm{km s}^{-1}$ relative to the LSR (optical velocity), and we iterated with a template that only contains subbands for which rest-frame CRRLs would appear at this redshift. 
This reduces the amount of noise which is cross-correlated, very slightly improving the definition of the peak of the cross-correlation. 

\begin{figure}
\begin{center}
\includegraphics[width=0.45\textwidth,clip, trim=0cm 0.5cm 0cm 0cm]{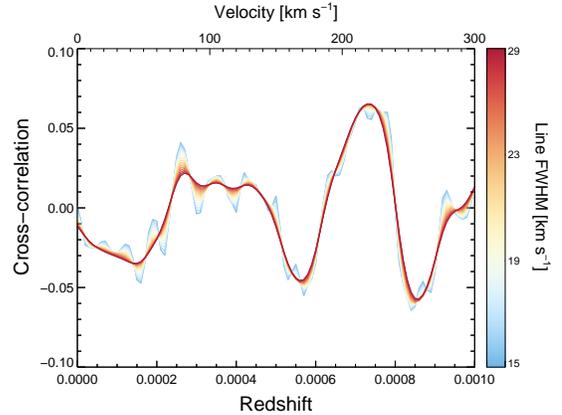}
\caption{Cross-correlation values versus redshift. Larger absolute values indicate a higher correlation between the template and observed spectra. \label{fig:f2}}
\end{center}
\end{figure}

Figure~\ref{fig:f2} shows a clear peak in the cross-correlation at $v=219\,\textrm{km s}^{-1}$ ($z=0.00073\pm0.00003$), consistent with the systemic velocity of M82, 
and we used this to stack our lines. 
We did not find evidence for CRRL features at either of the velocities of the two secondary peaks, $v=168\,\textrm{km s}^{-1}$ and $v=255\,\textrm{km s}^{-1}$. 

\subsection{Reconstructing the Line Profile}
\label{s33}
\begin{figure}
\begin{center}
\includegraphics[width=0.5\textwidth,clip, trim=2cm 4cm 1cm 1cm]{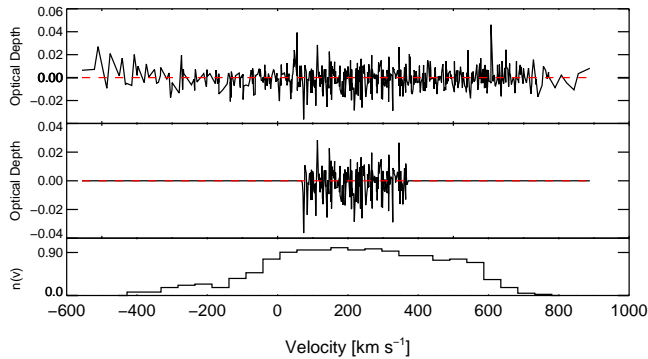}
\caption{The stacked spectrum with the expected CRRL feature at the measured systemic velocity of $v=219\,\textrm{km s}^{-1}$. From top to bottom: the complete reconstructed spectrum; the portion of the spectrum with dense velocity coverage; the normalised density of points in the spectrum. The red dashed line in the top two panels represents the continuum.\label{fig:f3}}
\end{center}
\end{figure}

Requiring that rest-frame CRRL frequencies be at least six channels away from the edge of a subband to avoid problems with noisy edge channels, we find that twenty-three subbands have $\alpha$-transitions. 
We clipped the first and last three edge channels and converted to optical velocity using the rest frequency of the CRRL within the subband. 
After blanking $\pm50\textrm{ km s}^{-1}$ around the expected velocity of the CRRL, we fitted the continuum in each subband individually with a low (first or second) order polynomial. 
We also tried blanking different ranges around the line, from $\pm50$ to $\pm250\textrm{ km s}^{-1}$, without seeing substantial differences in the final spectrum. 
For one subband, any blanking left continuum only on one side of the line, so we did not include this subband in the final stack. 
Each subband was continuum subtracted. 
The final spectrum was constructed from the individual points of 22 subbands, see Fig.~\ref{fig:f3}. 

The final stacked spectrum has a velocity sampling of $\sim1.5\,\textrm{km s}^{-1}$ within approximately $\pm150\,\textrm{km s}^{-1}$ of the measured velocity ($219\,\textrm{km s}^{-1}$, LSR). The measurement error of each point is equal to the standard deviation of the continuum in the subband from which the point originates. The weighted standard deviation of the continuum in the final spectrum within approximately $\pm150\,\textrm{km s}^{-1}$ of the expected CRRL line centre is $\sigma_{\tau}=0.005$, which is approximately $\sqrt{22}$ smaller than the average noise in a subband. 

\begin{figure}
\begin{center}
\includegraphics[width=0.5\textwidth,clip, trim=0.9cm 2.5cm 1cm 4cm]{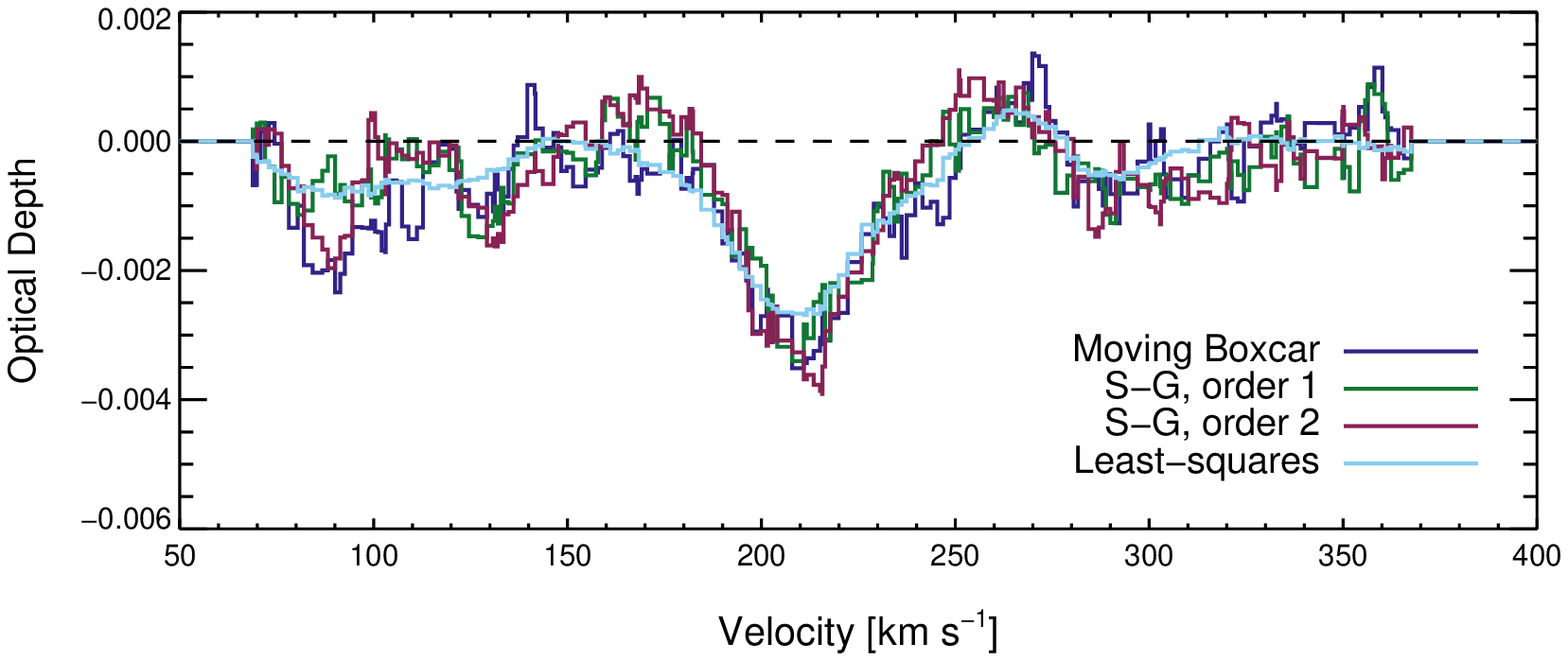}
\includegraphics[width=0.5\textwidth,clip, trim=1.8cm 6.5cm 1cm 1cm]{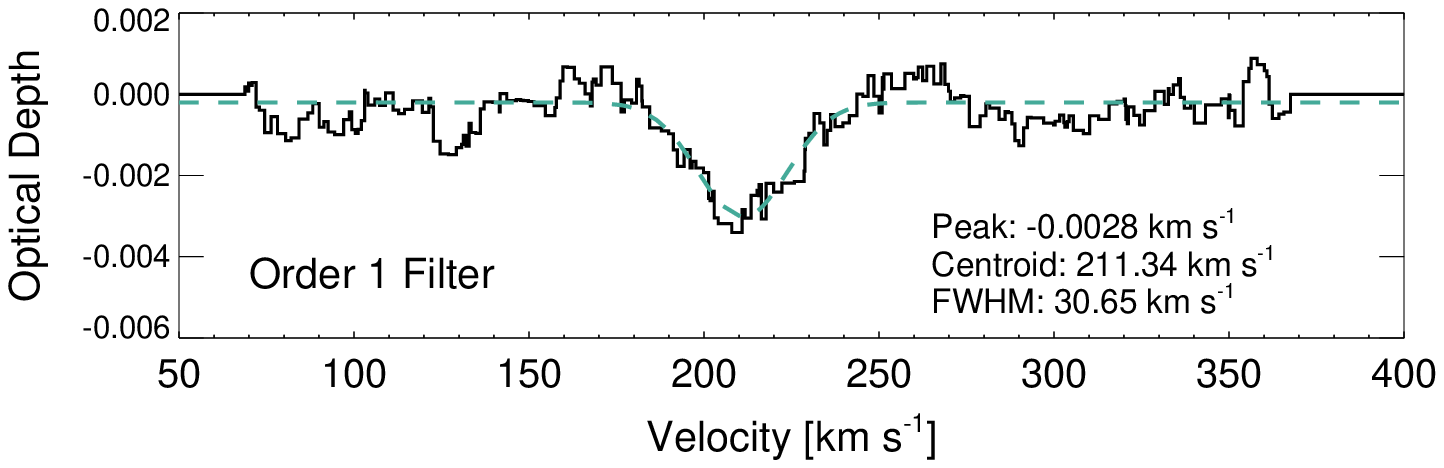}
\caption{Top: Different methods of smoothing the noise. The various methods provide similar line profiles. Bottom: The final spectrum, fitted with a Gaussian profile. \label{fig:f4}}
\end{center}%
\end{figure}

Galactic CRRLs show peak optical depths of $10^{-3}-10^{-4}$ \citep[e.g.,][]{ka01,ks81}. If the CRRLs in M82 have similar peak optical depths then these would be within the noise of this spectrum. We therefore used a low-pass Savitzky-Golay filter \citep[SGF;][]{sg64} 
which is a special case of a least-squares (LS) smoothing function that convolves the data with a filter whose shape is dependent on the polynomial order used for fitting. 
Other filtering methods (see Fig.~\ref{fig:f4}) produced similar line profiles. 
After testing SGF filter widths from 15 to 60 data points for both first and second order polynomials, we selected an SGF with a width of 31 data points, which provided the flattest continuum. 
Within approximately $\pm200\,\textrm{km s}^{-1}$ of the feature in the final, smoothed spectrum, the standard deviation of the continuum is $\sigma_{\tau}=3.3\times10^{-4}$, in good agreement with  our expectations from the raw spectrum noise and filter window size.
We assessed this filtering method with modelled spectra and found that the integrated line profile is preserved within the errors even after introducing Gaussian noise. 

Radial velocities within the nuclear region range from $\sim120-300\,\textrm{km s}^{-1}$. The centroid of the CRRL feature is therefore consistent with an absorption feature associated with the centre of the nuclear region. 
The cross-correlation is already strong evidence that the feature does not arise merely from a chance alignment of noise, since the feature only arises from a location where we know there to be gas, which is necessary but not sufficient for CRRLs. 
To rule out other possibilities, we conducted a series of tests. 
We used a `jack-knife' procedure where we re-stacked the subbands, each time leaving out an individual subband. 
None of the stacks produced a significant change to the absorption feature, confirming that the stacked spectrum is not dominated by one subband. 
Additionally, we stacked the subbands after introducing random velocity shifts to each subband before stacking, and were unable to produce any credible features. 
Spectra extracted from the background sky also do not show an absorption feature when stacked. 
From these tests, we conclude that our feature is real and associated with the M82 starburst galaxy.

We tried to detect CRRL $\beta$-transitions ($\Delta n=2$), without success. 
This is not unexpected, as the CRRL $\beta$-transitions are expected to have integrated optical depths that are only $15-30\%$ that of the $\alpha$-transitions \citep[e.g.,][]{stepkin07,pae89}. 
We did not find evidence for local CRRLs at $z=0$.
A search for hydrogen RRLs, both locally and at the redshift of M82, was also unproductive. This is unsurprising as these are believed to be very weak at these frequencies \citep[e.g., Salgado et al.  in preparation;][]{shaver75}. 

\section{Results}
We stacked the spectra to search for CRRLs in M82 using our measured velocity, $v=219\,\textrm{km s}^{-1}$, and detect carbon $\alpha$-transition RRLs with a combined signal-to-noise ratio of $8.5\sigma$ in the filtered spectrum. 
The central absorption feature can be fitted by a four parameter Gaussian profile with a depth of $2.8_{-0.10}^{+0.12}\times10^{-3}$, a FWHM of $30.6_{-1.0}^{+2.3}\textrm{ km s}^{-1}$, a centre of $211.3_{-0.5}^{+0.7}\textrm{ km s}^{-1}$, and an additive offset of $-2_{-0.072}^{+0.012}\times10^{-4}$ (see Fig.~\ref{fig:f4}).

\section{Discussion}
\label{sec:diss}
Other absorption features have been observed in M82 with similarly narrow widths. 
\citet{weiss10} observed p-H$_2$O absorption (in the far infrared) with a FWHM of $60\,\textrm{km s}^{-1}$, although this is offset from the CRRL absorption by $\sim50\,\textrm{km s}^{-1}$. 
Additionally, a narrow component results if we limit the reconstruction of the \ion{H}{1} absorption profile in the direction of SNRs in the nucleus of M82 to only the handful of bright SNRs which are relatively unaffected by free-free absorption at low frequencies. We constructed the \ion{H}{1} absorption line profiles from measurements in \citet{wills98}. We interpolated the spectra (mJy/bm) in the direction of each of the 26 SNRs which show \ion{H}{1} absorption onto a velocity grid with $1\,\textrm{km s}^{-1}$ resolution using a cubic spline function. Addition of the individual spectra produces a combined spectrum with information in the direction of SNRs. For the low frequency spectrum, we cross matched the SNRs in Table 1 of \citet{wills98} with those still seen by LOFAR at $154\,\textrm{MHz}$ (Varenius et al., in preparation). The eight SNRs from \citet{wills98} that have measured \ion{H}{1} absorption and are $>1\,\textrm{mJy}$ at 154 MHz are listed in that paper as 39.10+57.3, 39.40+56.1, 39.77+56.9, 42.53+61.9, 43.31+59.2, 45.74+65.2, 45.89+63.8, and 46.52+63.9. 
 
\begin{figure}
\begin{center}
\includegraphics[width=0.5\textwidth,clip, trim=0.9cm 7.5cm 1cm 1cm]{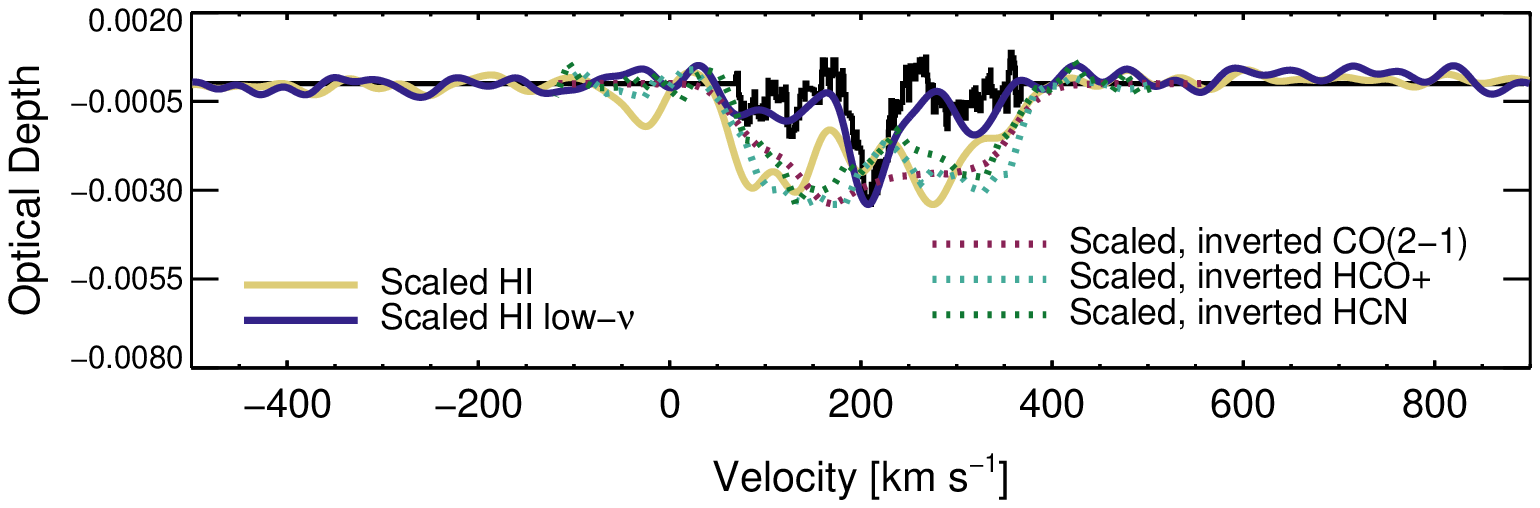}
\includegraphics[width=0.5\textwidth,clip, trim=0.9cm 7.5cm 1cm 1cm]{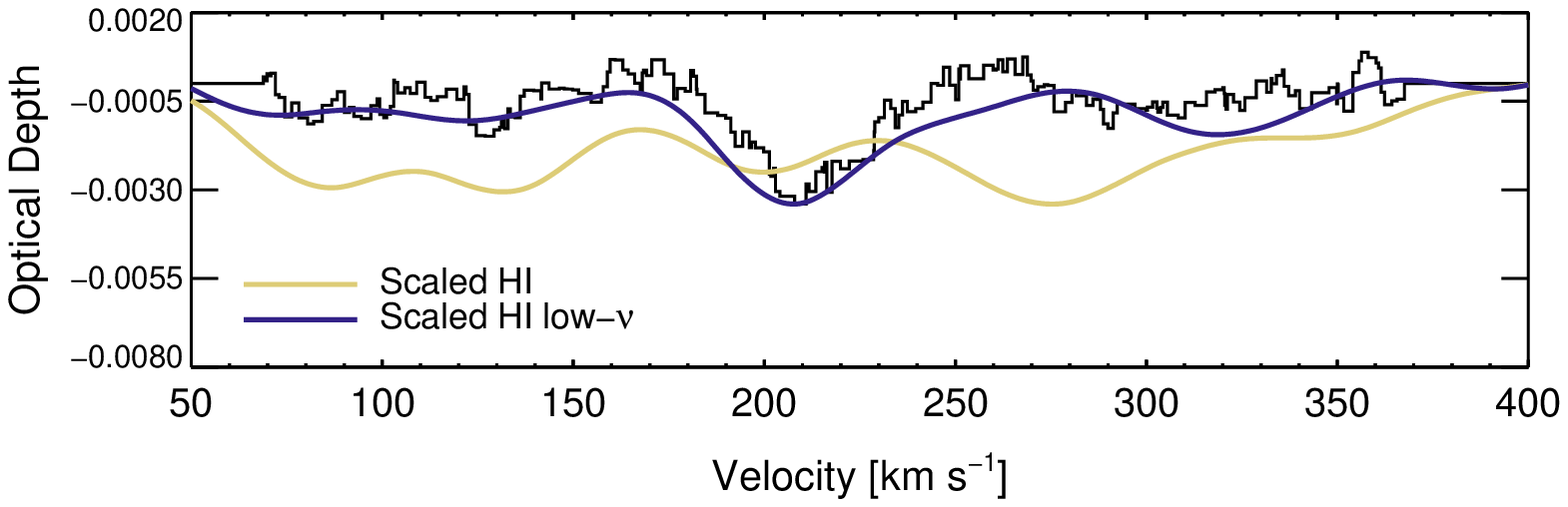}
\caption{Comparison with other tracers. The CRRL profile is plotted in black. Molecular emission lines \citep{kepley14} are shown by the dashed lines and are inverted for easier comparison, and \ion{H}{1} absorption in the direction of SNRs in M82 \citep{wills98} is plotted with solid colored lines. The gold solid line is the total contribution to the \ion{H}{1} absorption profile in the direction of all SNRs in M82, while the dark blue line is the contribution from only those SNRs relatively unaffected by free-free absorption at low frequencies. The bottom panel shows the central portion of the spectra, with only the \ion{H}{1} absorption profiles overlaid on the CRRL spectrum. 
\label{fig:f5}}
\end{center}
\end{figure}

Both of these absorption profiles show broader, shallower structure to either side of a deep, central feature, of which we see hints in the reconstructed CRRL profile (see Fig.~\ref{fig:f5}). 
Galactic observations show a correlation between CRRLs and \ion{H}{1} \citep[e.g.,][]{ka01,pae89}, so it is unsurprising that this correlation would also be present in M82. 

As Fig.~\ref{fig:f5} illustrates, integration of other tracers of interstellar gas over the disk of M82 leads to FWHM in excess of $200\,\textrm{ km s}^{-1}$ \citep{kepley14,wills98}. 
This leads us to expect a broad component of the CRRL profile, although the signal to noise coupled with the fixed velocity coverage and location of CRRLs in the subbands makes this difficult to ascertain. 
The overall CRRL profile follows the low-frequency \ion{H}{1} line profile (Fig.~\ref{fig:f5}). This shows that the CRRL absorbing gas is preferentially observed in front of those continuum sources that are still bright at low frequencies. The continuum offset and low-level absorption features in the CRRL spectrum also show reasonable agreement with the \ion{H}{1} absorbing gas, indicating that possibly we are observing a widespread CRRL component associated with cold, diffuse \ion{H}{1} throughout M82.

Integrating the line profile gives $\int\tau_{\textrm{CII}}\textrm{d}\nu=21.3\,\textrm{s}^{-1}$.  To constrain physical parameters of the gas, i.e. temperature and density, a single measurement is not enough. Additional measurements at higher frequencies (100-400 MHz) are necessary.

\section{Conclusions}
\label{sec:conclusions}
We have presented the first detection of extragalactic carbon radio recombination lines, in M82, using early science (Cycle 0) observations with LOFAR. While we already know that cold, neutral gas exists in M82 from \ion{H}{1} measurements, this is the first independent measurement of the atomic CNM. 
The narrow CRRL line is at a velocity corresponding to the centre of M82 and the profile corresponds to the absorption feature seen in the \ion{H}{1} spectrum. 
Higher resolution, higher sensitivity studies of M82 at the same frequencies would help to confirm that this feature is indeed associated with the nuclear region, and potentially shed more light on the origin of the optical depth. To constrain the gas properties we need studies at higher frequencies, and we are working towards detecting CRRLs in the range of $120-240\,\textrm{MHz}$ with LOFAR's high band antenna array. 

This discovery paves the way for future extragalactic CRRL studies to trace the CNM throughout the formation and evolution of galaxies, and is the basis of a pilot survey for CRRLs in other extragalactic sources. 

\section*{Acknowledgements}
LKM acknowledges financial support from NWO Top LOFAR project, project n. 614.001.006. JBRO acknowledges financial support from NWO Top LOFAR-CRRL project, project n. 614.001.351. The authors thank A. Kepley for providing GBT observed line profiles.


\begin{thebibliography}{99}

\bibitem[Adebahr et al.(2013)]{adebahr13} Adebahr, B., Krause, M., Klein, U., et al.\ 2013, \aap, 555, A23

\bibitem[Anantharamaiah et al.(1993)]{anath93} Anantharamaiah, K.~R., Zhao, J.-H., Goss, W.~M., \& Viallefond, F.\ 1993, \apj, 419, 585 

\bibitem[Anderson et al.(2011)]{anderson11} Anderson, L.~D., Bania, T.~M., Balser, D.~S., \& Rood, R.~T.\ 2011, \apjs, 194, 32 

\bibitem[Argo et al.(2010)]{argo10} Argo, M.~K., Pedlar, A., Beswick, R.~J., Muxlow, T.~W.~B., \& Fenech, D.~M.\ 2010, \mnras, 402, 2703

\bibitem[Asgekar et al.(2013)]{asgekar13} Asgekar, A., Oonk, J.~B.~R., Yatawatta, S., et al.\ 2013, \aap, 551, L11 

\bibitem[Briggs(1995)]{briggs95} Briggs, D.~S.\ 1995, Bulletin of the American Astronomical Society, 27, \#112.02 

\bibitem[Dupree(1969)]{dupree69} Dupree, A.~K.\ 1969, \apj, 158, 491 

\bibitem[Erickson et al.(1995)]{ema95} Erickson, W.~C., McConnell, D., \& Anantharamaiah, K.~R.\ 1995, \apj, 454, 125 

\bibitem[Fenech et al.(2010)]{fenech10} Fenech, D., Beswick, R., Muxlow, T.~W.~B., Pedlar, A., \& Argo, M.~K.\ 2010, \mnras, 408, 607 

\bibitem[Gandhi et al.(2011)]{gandhi11} Gandhi, P., Isobe, N., Birkinshaw, M., et al.\ 2011, \pasj, 63, 505 

\bibitem[Golynkin \& Konovalenko(1991)]{gk91} Golynkin, A.~A., \& Konovalenko, A.~A.\ 1991, Soviet Astronomy Letters, 17, 7

\bibitem[Jacobs et al.(2009)]{jacobs09} Jacobs, B.~A., Rizzi, L., Tully, R.~B., et al.\ 2009, \aj, 138, 332 

\bibitem[Kantharia \& Anantharamaiah(2001)]{ka01} Kantharia, N.~G., \& Anantharamaiah, K.~R.\ 2001, Journal of Astrophysics and Astronomy, 22, 51

\bibitem[Kepley et al.(2014)]{kepley14} Kepley, A.~A., Leroy, A.~K., Frayer, D., et al.\ 2014, \apjl, 780, L13 

\bibitem[Konovalenko \& Sodin(1981)]{ks81} Konovalenko, A.~A., \& Sodin, L.~G.\ 1981, \nat, 294, 135 

\bibitem[Konovalenko \& Stepkin(2005)]{ks05} Konovalenko, A.~A., \& Stepkin, S.~V.\ 2005, in Gurvits, L.~I., Frey, S., Rawlings, S., eds, EAS Publications Series, Vol. 15, Radio Recombination Lines. Cambridge Univ. Press, Cambridge, p. 271

\bibitem[Lo et al.(1987)]{lo87} Lo, K.~Y., Cheung, K.~W., Masson, C.~R., et al.\ 1987, \apj, 312, 574 

\bibitem[McDonald et al.(2002)]{mcd02} McDonald, A.~R., Muxlow, T.~W.~B., Wills, K.~A., Pedlar, A., 
\& Beswick, R.~J.\ 2002, \mnras, 334, 912 


\bibitem[Muxlow et al.(1994)]{muxlow94} Muxlow, T.~W.~B., Pedlar, A., Wilkinson, P.~N., et al.\ 1994, \mnras, 266, 455 

\bibitem[Offringa(2010)]{offringa10} Offringa, A.~R.\ 2010, Astrophysics Source Code Library, 10017 

\bibitem[Offringa et al.(2013)]{offringa13} Offringa, A.~R., de Bruyn, A.~G., Zaroubi, S., et al.\ 2013, \aap, 549, A11

\bibitem[Oonk et al.(2014)]{oonk14} Oonk, J.~B.~R., van Weeren, R.~J., Salgado, F., et al.\ 2014, \mnras, 437, 3506 

\bibitem[Palmer(1967)]{palmer67} Palmer, P.\ 1967, \apj, 149, 715 

\bibitem[Pandey et al.(2009)]{pandey09} Pandey, V.~N., van Zwieten, J.~E., de Bruyn, A.~G., \& Nijboer, R.\ 2009, The Low-Frequency Radio Universe, 407, 384 

\bibitem[Payne et al.(1989)]{pae89} Payne, H.~E., 
Anantharamaiah, K.~R., \& Erickson, W.~C.\ 1989, \apj, 341, 890 

\bibitem[Poppi et al.(2007)]{poppi07} Poppi, S., Tsivilev, A.~P., Cortiglioni, S., Palumbo, G.~G.~C., \& Sorochenko, R.~L.\ 2007, \aap, 464, 995

\bibitem[Rodriguez-Rico et al.(2004)]{rr04} Rodriguez-Rico, C.~A., Viallefond, F., Zhao, J.-H., Goss, W.~M., \& Anantharamaiah, K.~R.\ 2004, \apj, 616, 783 

\bibitem[Roelfsema \& Goss(1991)]{rg91} Roelfsema, P.~R., \& Goss, W.~M.\ 1991, \aaps, 87, 177 

\bibitem[Roelfsema et al.(1992)]{rgm92} Roelfsema, P.~R., Goss, W.~M., \& Mallik, D.~C.~V.\ 1992, \apj, 394, 188 

\bibitem[Roy et al.(2008)]{roy08} Roy, A.~L., Goss, W.~M., \& Anantharamaiah, K.~R.\ 2008, \aap, 483, 79 

\bibitem[Savitzky \& Golay(1964)]{sg64} Savitzky, A., \& Golay, M.~J.~E.\ 1964, Analytical Chemistry, 36, 1627 

\bibitem[Shaver(1975)]{shaver75} Shaver, P.~A.\ 1975, \aap, 43, 465 

\bibitem[Shaver et al.(1977)]{shaver77} Shaver, P.~A., Churchwell, E., \& Rots, A.~H.\ 1977, \aap, 55, 435 

\bibitem[Stepkin et al.(2007)]{stepkin07} Stepkin, S.~V., Konovalenko, A.~A., Kantharia, N.~G., \& Udaya Shankar, N.\ 2007, \mnras, 374, 852 

\bibitem[Tasse et al.(2013)]{tasse13} Tasse, C., van der Tol, S., van Zwieten, J., van Diepen, G., \& Bhatnagar, S.\ 2013, \aap, 553, A105 

\bibitem[van~Haarlem et al.(2013)]{vh13} van Haarlem, M.~P., Wise, M.~W., Gunst, A.~W., et al.\ 2013, \aap, 556, A2 

\bibitem[Walmsley \& Watson(1982)]{ww82} Walmsley, C.~M., \& Watson, W.~D.\ 1982, \apj, 260, 317

\bibitem[Wei{\ss} et al.(2010)]{weiss10} Wei{\ss}, A., Requena-Torres, M.A., G{\"u}sten, R., et al.\ 2010, \aap, 521, L1 

\bibitem[Westmoquette et al.(2013)]{westmoquette13} Westmoquette, M.~S., Smith, L.~J., Gallagher, J.~S., \& Walter, F.\ 2013, \mnras, 428, 1743 

\bibitem[Wills et al.(1998)]{wills98} Wills, K.~A., Pedlar, A., \& Muxlow, T.~W.~B.\ 1998, \mnras, 298, 347 

\bibitem[Wills et al.(1997)]{wills97} Wills, K.~A., Pedlar, A., Muxlow, T.~W.~B., \& Wilkinson, P.~N.\ 1997, \mnras, 291, 517 

\bibitem[Yun et al.(1993)]{yun93} Yun, M.~S., Ho, P.~T.~P., \& Lo, K.~Y.\ 1993, \apjl, 411, L17 

\end{thebibliography}
\end{document}